\documentclass[modern,trackchanges]{aastex62}

\newcommand\rh{\ensuremath{R_{\mathrm{h}}}}
\newcommand\coo{CO\ensuremath{_2}}

\usepackage{color,soul}

\definecolor{darkgreen}{rgb}{0,0.7,0.0}

\submitjournal{ApJ Letters}

\shortauthors{Protopapa et al.}

\begin{document}

\title{Icy Grains from the Nucleus of Comet C/2013 US$_{10}$ (Catalina)}

\author[0000-0001-8541-8550]{Silvia Protopapa}
\altaffiliation{Visiting Astronomer at the Infrared Telescope Facility,\\ which is operated by the University of Hawaii under contract\\ NNH14CK55B with the National Aeronautics and\\ Space Administration.}
\affiliation{Department of Astronomy, University of Maryland, College Park, MD, USA}
\affiliation{Southwest Research Institute, Boulder, CO 80302, USA}

\author[0000-0002-6702-7676]{Michael S. P. Kelley}
\altaffiliation{Visiting Astronomer at the Infrared Telescope Facility,\\ which is operated by the University of Hawaii under contract\\ NNH14CK55B with the National Aeronautics and\\ Space Administration.}
\affiliation{Department of Astronomy, University of Maryland, College Park, MD, USA}

\author{Bin Yang}
\altaffiliation{Visiting Astronomer at the Infrared Telescope Facility,\\ which is operated by the University of Hawaii under contract\\ NNH14CK55B with the National Aeronautics and\\ Space Administration.}
\affiliation{European Southern Observatory, Santiago, Chile}

\author[0000-0001-9542-0953]{James M. Bauer}
\affiliation{Department of Astronomy, University of Maryland, College Park, MD, USA}

\author{Ludmilla Kolokolova}
\affiliation{Department of Astronomy, University of Maryland, College Park, MD, USA}

\author{Charles E. Woodward}
\altaffiliation{Visiting Astronomer at the Infrared Telescope Facility,\\ which is operated by the University of Hawaii under contract\\ NNH14CK55B with the National Aeronautics and\\ Space Administration.}
\affiliation{Minnesota Institute for Astrophysics, University of Minnesota, Minneapolis, MN 55455, USA}

\author[0000-0002-2021-1863]{Jacqueline V. Keane}
\affiliation{Institute for Astronomy, 2680 Woodlawn Drive, Honolulu, HI 96822, USA}

\author[0000-0002-9413-8785]{Jessica M. Sunshine}
\affiliation{Department of Astronomy, University of Maryland, College Park, MD, USA}

\correspondingauthor{Silvia Protopapa}
\email{sprotopapa@boulder.swri.edu}



\begin{abstract}
We present IRTF/SpeX and \textit{NEOWISE} observations of the dynamically new comet C/2013 US$_{10}$ (Catalina), hereafter US10, from 5.8~au inbound, to near perihelion at 1.3~au, and back to 5.0~au outbound. We detect water ice in the coma of US10, assess and monitor the physical properties of the ice as insolation varies with heliocentric distance, and investigate the relationship between water ice and \coo. This set of measurements is unique in orbital coverage and can be used to infer the physical evolution of the ice and, potentially, the nucleus composition. We report (1) nearly identical near-infrared spectroscopic measurements of the coma at $-$5.8~au, $-$5.0~au, +3.9~au (where $<$0~au indicates pre-perihelion epochs), all presenting evidence of water-ice grains, (2) a dust-dominated coma at 1.3~au and 2.3~au and, (3) an increasing \coo/$Af\rho$ ratio from $-$4.9~au to 1.8~au. We propose that sublimation of the hyper-volatile \coo~is responsible for dragging water-ice grains into the coma throughout the orbit. Once in the coma, the observability of the water-ice grains is controlled by the ice grain sublimation lifetime, which seems to require some small dust contaminant (i.e., non-pure ice grains). At $|\rh|\geq3.9$~au, the ice grains are long-lived and may be unchanged since leaving the comet nucleus. We find the nucleus of comet US10 is made of, among other components, $\sim$1-\micron~water-ice grains containing up to 1\% refractory materials.
\end{abstract}



\section{Introduction} \label{sec:intro}
The structure of cometary nuclei is still a subject of debate. It is predicted by several models that consider different origins and/or evolutionary processes \citep[e.g.,][]{Greenberg1999, Davidsson2016}. Investigation of the physical properties of the major constituent water ice, such as phase, purity, particle size, are required to validate or disprove such models. As an example, the presence of pure, fine grained water-ice particles in the impact ejecta of comet 9P/Tempel 1 led \citet{Sunshine2007} to question the comet formation model of \citet{Greenberg1999}. It is difficult to directly study the properties of comet nuclei. Instead, we can study comet interiors through their comae, as cometary activity, i.e., sublimation-driven mass loss, makes the primitive interiors of these objects available for characterization. However, we must determine how coma material evolves after leaving the nucleus before drawing conclusions about cometary interiors, and, ultimately, how they formed. This is especially true for volatile materials, such as water-ice grains. By observing water ice in comae, and monitoring its properties over time and a wide range of heliocentric distances, we can infer its physical evolution in the coma and, potentially, the nucleus composition.

Besides water ice, comets contain a plethora of more volatile materials \citep[e.g.,][]{Cochran2015, LeRoy2015}. In the typical comet, \coo~is the second most abundant volatile, with coma mixing ratios near 5 to 25\% that of water \citep{Ootsubo2012, AHearn2012}. In the classical comet model, water ice controls sublimation and the mass-loss process. However, the Deep Impact eXtended Investigation (DIXI) flyby of comet 103P/Hartley~2 provided new insight into the critical role of \coo~as a driver of the activity \citep{AHearn2011}. The close-up view of 103P/Hartley~2 revealed bright jets rich with water-ice grains in the inner few km of the coma, driven off the nucleus by the sublimation of the hyper-volatile \coo~\citep{Protopapa2014}. This \coo-driven process occurred near 1.0~au from the Sun, despite the greater abundance and high sublimation rate of water ice at this distance. The DIXI observations have formed the basis for our current understanding of the presence of water-ice grains in comet comae. However, the relationship between the \coo~production rate and the presence of a water-ice grain halo has yet to be explored in other comets. \coo~has been mapped in the coma of 9P/Tempel 1 \citep{Feaga2007} and 67P/Churyumov-Gerasimenko \citep{Fink2016} as seen by the Deep Impact and Rosetta missions, respectively. However, no water-ice grains have been reported in the quiescent coma of both comets. On the other hand, 103P/Hartley~2 is known as an hyperactive comet \citep{Groussin2004}, a class of comets with an extremely high water production rate for the given nucleus size.  The presence of water ice in the coma may account for the high production rate, but a full characterization of the water sources is complex \citep{AHearn2011, Bonev2013, Kelley2013, Kelley2015, Knight2013, Protopapa2014, Belton2017}.

Here, we present IRTF/SpeX and \textit{NEOWISE} observations of the dynamically new comet C/2013 US$_{10}$ (Catalina), hereafter US10, which was discovered on 31 October 2013 by the Catalina Sky Survey at 8~au \citep{Honkova2013} and reached a perihelion distance of 0.82~au on 15.7 November 2015. Observations of US10 were acquired pre- and post-perihelion over a broad range of heliocentric distances (\rh{}). The main goals of this study are to (1) detect water-ice grains, (2) assess and monitor the physical properties of the ice as insolation varies with heliocentric distance, and (3) investigate the relation between water ice and \coo. 

\section{Observations} \label{sec:obs}

\begin{deluxetable*}{ccccccccccccc}
\tablecaption{Observational Parameters for C/2013 US10 (Catalina).\label{tab_obs}}
\tabletypesize{\tiny}
\tablecolumns{14}
\tablewidth{0pt}
\tablehead{
\colhead{UT date} &
\colhead{Mode} & 
\colhead{Slit\tablenotemark{a}} & 
\colhead{AM$_{target}$\tablenotemark{b}} & 
\colhead{SA (spectral type)\tablenotemark{c}} &
\colhead{AM$_{SA}$\tablenotemark{d}} & 
\colhead{\rh{}\tablenotemark{e}} &
\colhead{$\Delta$\tablenotemark{f}} & 
\colhead{T-mag\tablenotemark{g}} &
\colhead{$F_\nu(3.4)$\tablenotemark{h}} &
\colhead{$F_\nu(4.6)$\tablenotemark{h}} & 
\colhead{$\log_{10}(Q_{\coo})$\tablenotemark{i}} & 
\colhead{$\log_{10}(Af\rho)$\tablenotemark{l}} \\
\colhead{(YYYY-MM-DD)} & 
\colhead{} &
\colhead{(arcsec)} & 
\colhead{} & 
\colhead{} &
\colhead{} &
\colhead{au} &
\colhead{au} &
\colhead{} &
\colhead{(mJy)} &
\colhead{(mJy)} &
\colhead{(s$^{-1}$)} &
\colhead{(cm)} \\
}
\startdata
\hline
IRTF & & & & & & &  & \\
\hline
2014-08-13 & Prism & 0.8$\times$15 & 1.27 & HD 222361 (G2V C)       & 1.29 & -5.8 & 5.0 & 17.0 & \nodata & \nodata & \nodata & \nodata \\
2014-11-08 & Prism & 0.8$\times$15 & 1.43 & HD 209548 (G2V C)       & 1.41 & -5.0 & 4.7 & 16.3 & \nodata & \nodata & \nodata & \nodata \\
2016-01-12 & Prism & 3.0$\times$60 & 1.14 & HD 106965 (A0V/1V D)    & 1.07 & 1.3 & 0.7 & 8.3 & \nodata & \nodata & \nodata & \nodata \\
2016-01-12 & LXD   & 0.8$\times$60 & 1.20 & HD 121880 (A0V D)       & 1.22 & 1.3 & 0.7 & 8.3 & \nodata & \nodata & \nodata & \nodata \\
2016-03-27 & Prism & 0.8$\times$15 & 1.53 & BD+45 978 (G2V C)       & 1.49 & 2.3 & 2.5 & 12.6 & \nodata & \nodata & \nodata & \nodata \\
2016-08-13 & Prism & 1.6$\times$60 & 1.81 & GSC 01881-01236 (G2V D) & 1.65 & 3.9 & 4.4 & 15.5 & \nodata & \nodata & \nodata & \nodata \\
\hline
\textit{NEOWISE} & & & & & & &  & \\
\hline
2014-06-21 & \nodata & \nodata & \nodata & \nodata & \nodata & -6.3 & 6.3 & \nodata & 0.21$\pm$0.05 & 0.14$\pm$0.04 & $<$26.52 & 3.0$\pm$0.1 \\
2014-11-19 & \nodata & \nodata & \nodata & \nodata & \nodata & -4.9 & 4.7 & \nodata & 0.75$\pm$0.17 & 1.04$\pm$0.23 & 26.95$^{+0.2}_{-0.3}$ & 3.2$\pm$0.1 \\
2015-06-07 & \nodata & \nodata & \nodata & \nodata & \nodata & -2.7 & 2.5 & \nodata & 12.3$\pm$2.7 & 22.5$\pm$4.9 & 27.66$^{+0.1}_{-0.2}$ & 3.6$\pm$0.1 \\
2015-08-28 & \nodata & \nodata & \nodata & \nodata & \nodata & -1.6 & 1.2 & \nodata & 80$\pm$18 & 323$\pm$71 & 27.76$^{+0.2}_{-0.3}$ & 3.6$\pm$0.1 \\
2016-01-08 & \nodata & \nodata & \nodata & \nodata & \nodata & 1.3 & 0.8 & \nodata & 234$\pm$51 & 1540$\pm$340 & $<$28.6 & 3.7$\pm$0.1 \\
2016-02-21 & \nodata & \nodata & \nodata & \nodata & \nodata & 1.8 & 1.4 & \nodata & 30$\pm$7 & 96$\pm$20 & 27.69$^{+0.2}_{-0.3}$ & 3.4$\pm$0.1\\
\enddata
\tablenotemark{a}{Slit width $\times$ slit length.}
\tablenotemark{b}{Average airmass of the comet observations.}
\tablenotemark{c}{Star used for calibration purposes, its spectral type and precision from SIMBAD.}
\tablenotemark{d}{Average airmass of the solar analog observations.}
\tablenotemark{e}{The Sun-to-target distance ($<$0~au indicates pre-perihelion epochs).}
\tablenotemark{f}{The target-to-observer distance.}
\tablenotemark{g}{Comet's approximate apparent visual total magnitude as reported by Horizons.}
\tablenotemark{h}{The spectral flux density at 3.6 and 4.6 $\mu$m.}
\tablenotemark{i}{\coo~production rate.}
\tablenotemark{l}{Product of the albedo ($A$), the filling factor ($f$), and the radius ($\rho$) of the coma.}
\end{deluxetable*}

\begin{figure}
\gridline{\fig{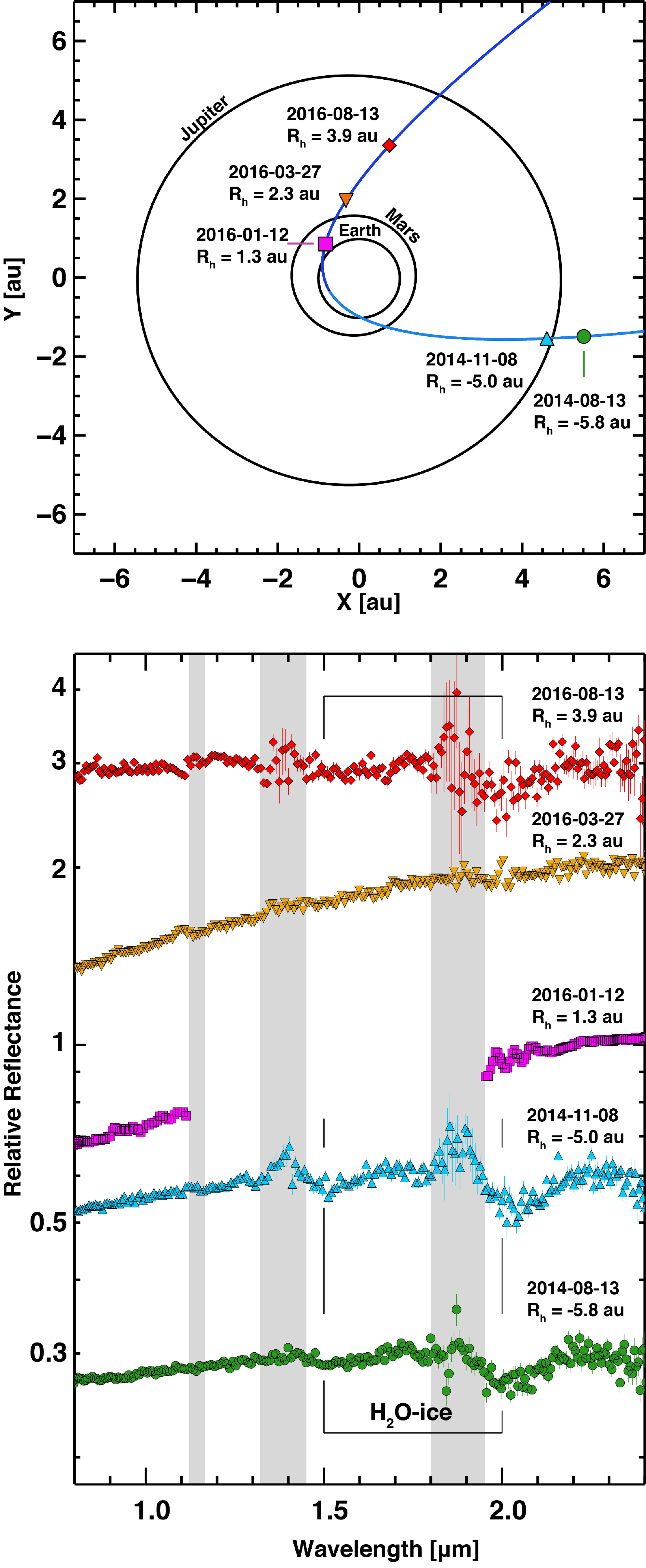}{0.45\textwidth}{}}
\caption{Temporal evolution of the coma properties of comet  C/2013 US10 (Catalina). \textit{Top panel}: Orbit and position of the comet during our observing campaign. \textit{Bottom panel}: IRTF/SpeX spectra of comet US10, binned to $R\sim250$. Spectra are normalized at 2.2~\micron{} to the values 0.3, 0.6, 1.0, 2.0, 3.0, in order from bottom to top. Gray shaded rectangles are regions of strong telluric absorption.\label{fig:obs}}
\end{figure}

Near-IR spectroscopy is an excellent tool for investigating the physical properties of water ice through its characteristic absorption bands at 1.5, 2.0, and 3.0~\micron{} \citep{Warren2008}. Near-IR spectra of US10 were acquired with the SpeX spectrograph \citep{Rayner2003} on the 3-m NASA Infrared Telescope Facility (IRTF) at Mauna Kea Observatory. Most observations were obtained using the high-throughput low-resolution prism mode covering the wavelength range 0.7--2.52~\micron{} (Figure \ref{fig:obs}). Slit widths of 0\farcs8, 1\farcs6, and 3\farcs0 were employed throughout our US10 observing campaign for a resulting spectral resolving power ($R\equiv\lambda/\Delta\lambda$) of 82, 41, and 20, respectively (Table \ref{tab_obs}). Given that water-ice absorption bands are broad, low resolution spectroscopy is sufficient for our purposes. On 2016 January 12, observations were also carried out with the long wavelength cross-dispersed (LXD) mode covering the wavelength range 1.67--4.2~\micron{} ($R\sim940$ for the 0\farcs8~slit).

Observations of US10 were obtained with the 15\arcsec{} and 60\arcsec{} long slits. We reduced the data using IDL and Python codes, based on Spextool \citep{Cushing2004}. The latter does not reduce observations taken with the 60\arcsec{} slit. Spectra were extracted with a $\sim$3\arcsec~aperture radius, then wavelength calibrated and combined using a robust mean, with a sigma clipping threshold of 2.5. The uncertainty is given by the standard deviation on the good pixels.

For most nights, spectra of nearby solar analog stars (G2V) were obtained in order to simultaneously remove the telluric and solar features from the comet spectra. The January 2016 LXD and prism data were instead calibrated with a nearby A0V star \citep[following][]{Vacca2003} and normalized with a solar spectrum.

\textit{NEOWISE} \citep{Mainzer2011} observed comet US10 over six visits between 2014 and 2016 with two broadband filters centered at 3.4 and 4.6~\micron.  Following \citet{Reach2013} and \citet{Bauer2015}, we assume that the 3.4~\micron{} image is dominated by dust, and the 4.6~\micron{} image is a combination of dust and gas emission.  For most epochs, we measured photometry of the comet using an 11\arcsec{} aperture radius. The January 2016 data was saturated in the inner 1.5\arcsec{} aperture radius, and instead we used an annulus with radii from 6\arcsec{} to 7\arcsec{} and scaled the photometry to account for the missing interior flux \citep{Jewitt1984}.  

\section{Results} \label{sec:results}
\subsection{Near-infrared spectroscopy}
\begin{figure*}
\plotone{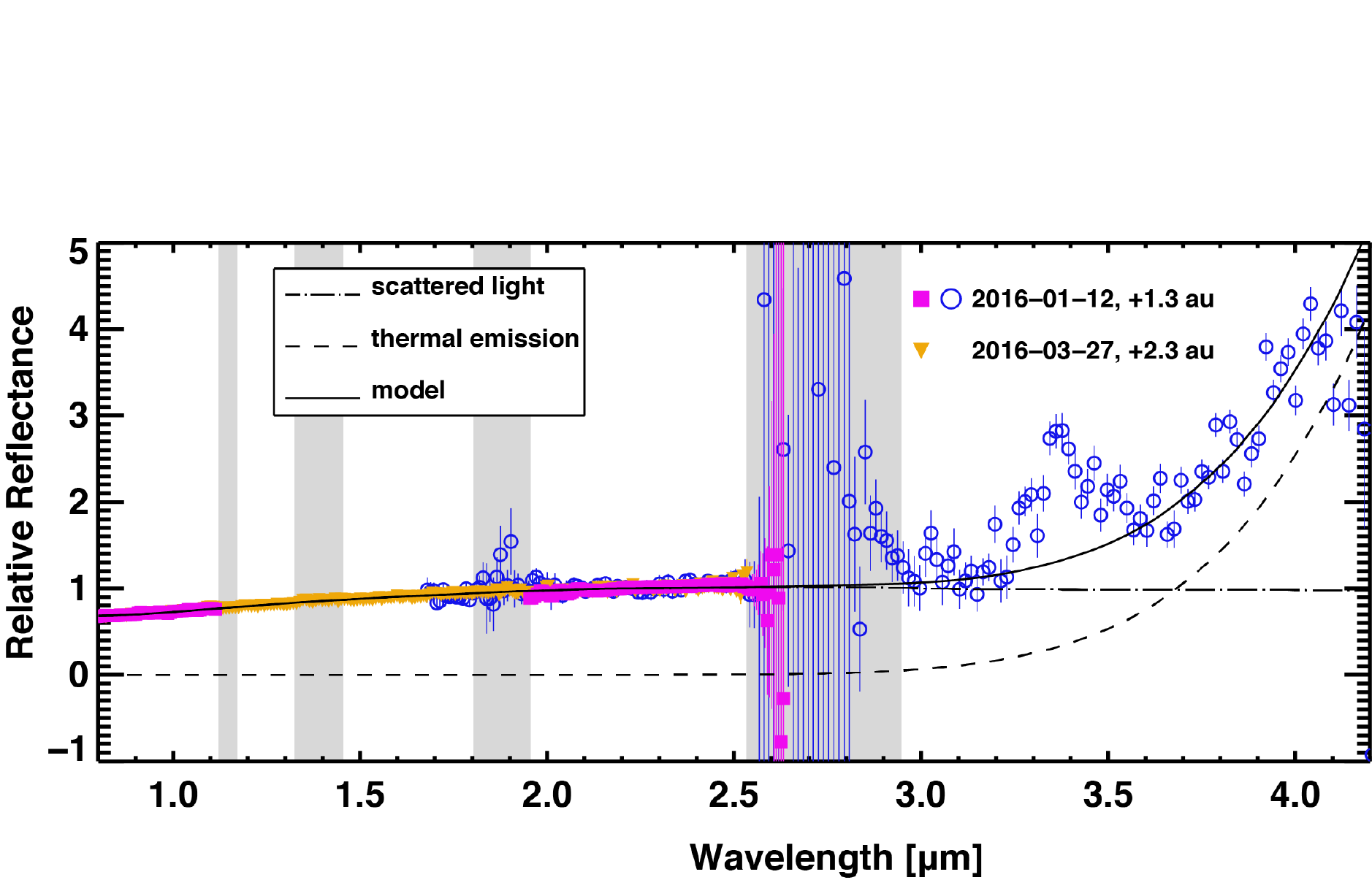}
\caption{Comet US10 reflectance spectra acquired close to perihelion over the wavelength range 0.8--4.2~\micron. The model (solid line) is the sum of the scattered solar radiation (dash-dot line) and a blackbody function (dashed line). The synthetic spectrum matches all of the main features of the observed spectrum. The organic emission feature near 3.3~\micron{} is not modeled. The gray shaded rectangles are regions of strong telluric absorption. The IRTF data behind this figure are available as machine-readable tables.\label{fig:obs_refractory}}
\end{figure*}
\begin{figure*}
\plotone{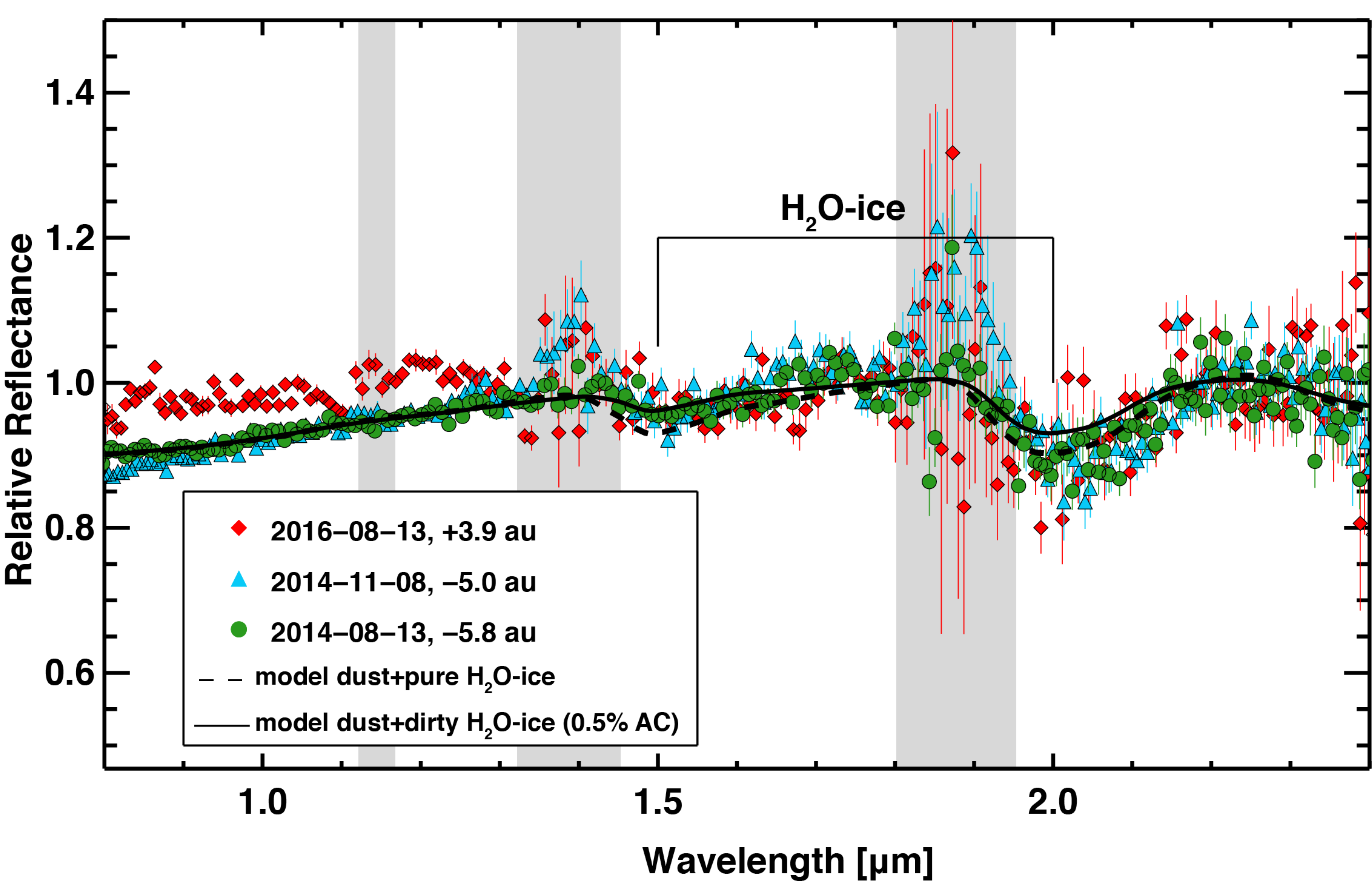}
\caption{IRTF/SpeX prism mode observations of US10 acquired at $|\rh|\geq$3.9~au. The reflectance spectra display water-ice absorption bands at 1.5 and 2.0~\micron. The shape and strength of the water-ice absorption features is consistent among the three data sets. The best-fit models from Sec.~\ref{sec:results} and Sec.~\ref{sec:discussion} are overplotted in dashed and solid line, respectively. The IRTF data behind this figure are available as machine-readable tables.\label{fig:obs_water}}
\end{figure*}

The spectra obtained close to perihelion at 1.3 and 2.3~au are nearly identical (Fig. \ref{fig:obs_refractory}). They are featureless and red sloped, which suggests a dust-dominated coma. The prism spectrum acquired at 1.3~au (magenta filled squares in Fig.~\ref{fig:obs} and \ref{fig:obs_refractory}) includes a gap in the range between 1.12 and 1.95~\micron{} due to artifacts. The LXD spectrum observed the same date is consistent with the prism measurements between 1.7 and 2.4~\micron, giving confidence to the short wavelength data. Beyond 2.5~\micron, the spectrum displays molecular/organic emission at 3.3--3.5~\micron{}, and a sharp rise at $>$3.5~\micron{} due to thermal emission from dust. The spectra do not present any significant absorption bands, not even the 3-\micron{} water-ice band, which has the largest absorption coefficient for water ice (by a factor of $\sim$10$^{2}$) and is therefore detectable even at low abundances. 

We model the featureless continuum by simultaneously fitting the scattered and thermal components similar to \citet{Protopapa2014}.  Instead of using a Hapke scattering model, we use Mie theory so that we can model grains outside the geometric optics regime. Use of a single particle size poorly reproduces the shape of the reflected continuum.  Therefore, we proceeded by considering a power-law size distribution of porous aggregates. Given the absence of water-ice absorptions, we limit the composition to a single refractory material. The average single scattering albedo is given by
\begin{equation}
\textrm{w} = \frac{\int_{a_{1}}^{a_{2}}\,a^2Q_{S}(a)n(a) da}{\int_{a_{1}}^{a_{2}}\,a^2Q_{E}(a)n(a) da}
\end{equation}
where $a_{1}$ and $a_{2}$ are the smallest and largest particle diameters in the size distribution, respectively, $Q_{S}(a)$ and $Q_{E}(a)$ are the scattering and extinction efficiencies of aggregates as a function of diameter $a$, respectively, and $n(a)da$ is the number of particles per unit volume with diameter between $a$ and $a+da$, following the formula
\begin{equation}
n(a) \propto{} a^{\alpha}
\end{equation}
with $a$ spanning the range 1--100~$\mu$m. We compute the scattering and extinction efficiencies by means of Mie scattering theory. The optical constants of the aggregate ($n(\lambda)$,$k(\lambda)$) are estimated from effective medium theory \citep[Bruggeman mixing formula,][]{Bohren1983}, where the effective medium is a mixture of vacuum and amorphous carbon \citep{Rouleau1991}. We assume the thermal emission can be described with a single Planck function in the limited wavelength range we are considering. The free parameters in our model are the exponent of the particle size distribution, $\alpha$, the volume fraction of the amorphous carbon in the grain, $f_{AC}$, and the coma temperature $T_{c}$. They are iteratively modified by means of a Levenberg-Marquardt $\chi^{2}$ minimization algorithm until the best fit to the observations is achieved. The best-fit model (Fig.~\ref{fig:obs_refractory}) has medium porosity dust grains $f_{AC}$=76.5$\pm$0.2\% with a differential dust size distribution power-law index of -4.63$\pm$0.01 and a temperature of 286$\pm$5~K. 

The spectra acquired at $|\rh|\geq 3.9$~au (Fig.~\ref{fig:obs_water}) pre- and post-perihelion display water-ice absorption bands at 1.5- and 2.0-$\mu$m. The spectra are in units of relative reflectance. The relative strength and shape of the 1.5- and 2.0-$\mu$m features depend on the grain size, presence of impurities, and the relative abundance of ice and dust in the coma. The three spectra shown in Fig.~\ref{fig:obs_water} are consistent in terms of relative strength and shape of the water-ice absorptions bands (variations in the slope at short wavelengths is likely related to the choice of solar analogs). This consistency suggests that when water ice is present, its properties are independent of heliocentric distance. This is a striking result as it suggests that the observed grains, once in the coma, do not experience major chemical and physical evolution in our field-of-view. The implication of this finding is discussed in the next section.

We model the $|\rh|\geq 3.9$~au spectra with two components, one of polydisperse porous dust particles, following the same approach as the featureless spectra, and one of polydisperse water-ice aggregates. To limit the number of free parameters, we hold fixed the dust grains using the best-fit values from the ice-free spectra and solve only for the ice-to-dust ratio and the properties of the water-ice aggregates. This initial fit yields an ice particle size distribution with a very steep slope, $\alpha\sim-14$. Therefore we proceeded to model the data assuming monodisperse water-ice particles. The revised best-fit model (dashed black line in Fig.~\ref{fig:obs_water}) suggests the presence of solid water-ice grains ($f_{H_{2}O}$=100\%) with a diameter of 1.2$\pm$0.2~\micron, and a coma ice-to-dust areal ratio of 15$\pm$1\% (reduced-$\chi^2$=1.4). The presented models are characterized by geometric albedos of 9\% and 17\% at 0.8~\micron{}, for the ice-free and icy comae, respectively, indicating they are realistic models \citep{Kolokolova2004}.  The icy coma model will be further refined after considering the grain lifetimes in Section~\ref{sec:discussion}.

\subsection{\coo{} and dust production}
Using the dust spectral model of \citet{Bauer2015}, we scale the 3.4-\micron{} photometry to an effective 4.6-\micron{} value, and subtract it from the 4.6-\micron{} photometry. We attribute any residual flux to gas emission in the band. Between the three major species with emission lines in the 4.6-\micron{} band (H$_2$O, \coo, and CO), \coo{} is the brightest for most comets \citep{Ootsubo2012}.  The gas flux is converted to \coo{} production rates following the methods of \citet{Stevenson2015} and \citet{Bauer2015}. The spectral model also provides an estimate of the cometary dust $Af\rho$ quantity, a value that is roughly proportional to dust production rate under some conditions and assumptions \citep{AHearn1984, Fink2015}. The flux values at 3.4 and 4.6~\micron, as well as the \coo{} and dust production rates determined at each epoch are reported in Table \ref{tab_obs}.

\section{Discussion}\label{sec:discussion}
Our spectra of comet US10 span a wide range of heliocentric distances, from 5.8~au pre-perihelion, to near perihelion at 1.3~au, and back out to 5.0~au post-perihelion. The only prior study investigating temporal changes of water-ice grains is that by \citet{Yang2014}, with pre-perihelion measurements of C/2011 L4 (Pan-STARRS) between 3.9 and 5.2~au. Therefore, our set of measurements is unique in orbital coverage and can be used to infer the physical evolution of the ice and, potentially, the nucleus composition.

The spectra obtained at $|\rh|\leq2.3$~au are dominated by dust. The modeling analysis proposed here indicates the coma is composed of aggregates of refractories spanning 1 to 100~\micron{} in diameter with a power-law size distribution. In contrast, the spectra acquired at $|\rh|\geq3.9$~au clearly exhibit strong water-ice features. Additionally, all the spectra with water-ice absorptions are nearly identical and consistent with the presence of solid water-ice particles on the order of 1~\micron{} in diameter spatially separated from the dust. 

The persistent presence of water-ice grains with similar physical properties at large heliocentric distances pre- and post-perihelion implies a lack of a significant grain physical and chemical evolution.  The changing ice fraction of the coma can be interpreted in terms of the limited lifetime of the water-ice grains with respect to the field-of-view when closer to the Sun. Grains near the Sun have higher sublimation rates and therefore shorter lifetimes.  Thus, the icy coma is limited in size.

We calculate the lifetimes of the water-ice grains in the coma of US10 with the goal of testing the idea that the same grains are ejected throughout the orbit and that sublimation is the main process responsible for the observed differences between our spectroscopic measurements. This is done by balancing the absorption of sunlight with energy lost through thermal emission and sublimation. We also include losses due to sputtering by solar wind \citep{Mukai1981}, which limits grain lifetimes when sublimation lifetimes are very long.  We find that pure water-ice grains with a 1.2-\micron{} diameter have a lifetime of $3\times10^{4}$ to $6\times10^{11}$~s at 1.3 to 5.8~au, in agreement with previous estimates \citep{Hanner1981, Lien1990, Beer2006}. For an assumed coma expansion speed of 10--100~m~s$^{-1}$, the lifetime at 2.3~au ($4\times10^8$~s) corresponds to a length-scale of $10^6$ to $10^7$~km.  This long lifetime is inconsistent with the absence of water ice at 2.3~au within a 1500-km slit width. However, if the water ice has a small amount of low albedo dust, the lifetimes will be significantly shorter. Assuming this is the case, the detection of ice at 3.9~au and the non-detection at 2.3~au can be used to estimate an approximate dust fraction.  With just 0.5\% amorphous carbon by volume mixed into the ice using effective medium theory, the lifetimes are reduced to $6\times10^3$ and $3\times10^5$~s, and the halo sizes to 60--600~km and 3\,000--30\,000~km at 2.3 and 3.9~au, respectively. Thus, a small dust fraction is consistent with our observations with 1500- and 5000-km slits at 2.3 and 3.9~au, respectively.

We refit our distant spectroscopy with 1.2-\micron{} solid-ice grains containing 0.5\% amorphous carbon by fraction.  The ice grain reflectance model is the same, except the absorption bands are decreased in strength by 3\%. The retrieved parameters are essentially unchanged: the revised coma ice fraction is 18.3$\pm$0.1\% and the reduced-$\chi^2$ is unchanged (solid black line in Fig.~\ref{fig:obs_water}).  We also tested a 1\% dirt fraction model. The results were acceptable for the lifetimes, but the spectral model was significantly changed and the reduced-$\chi^2$ increased from 1.4 to 1.7.

\begin{figure}
\plotone{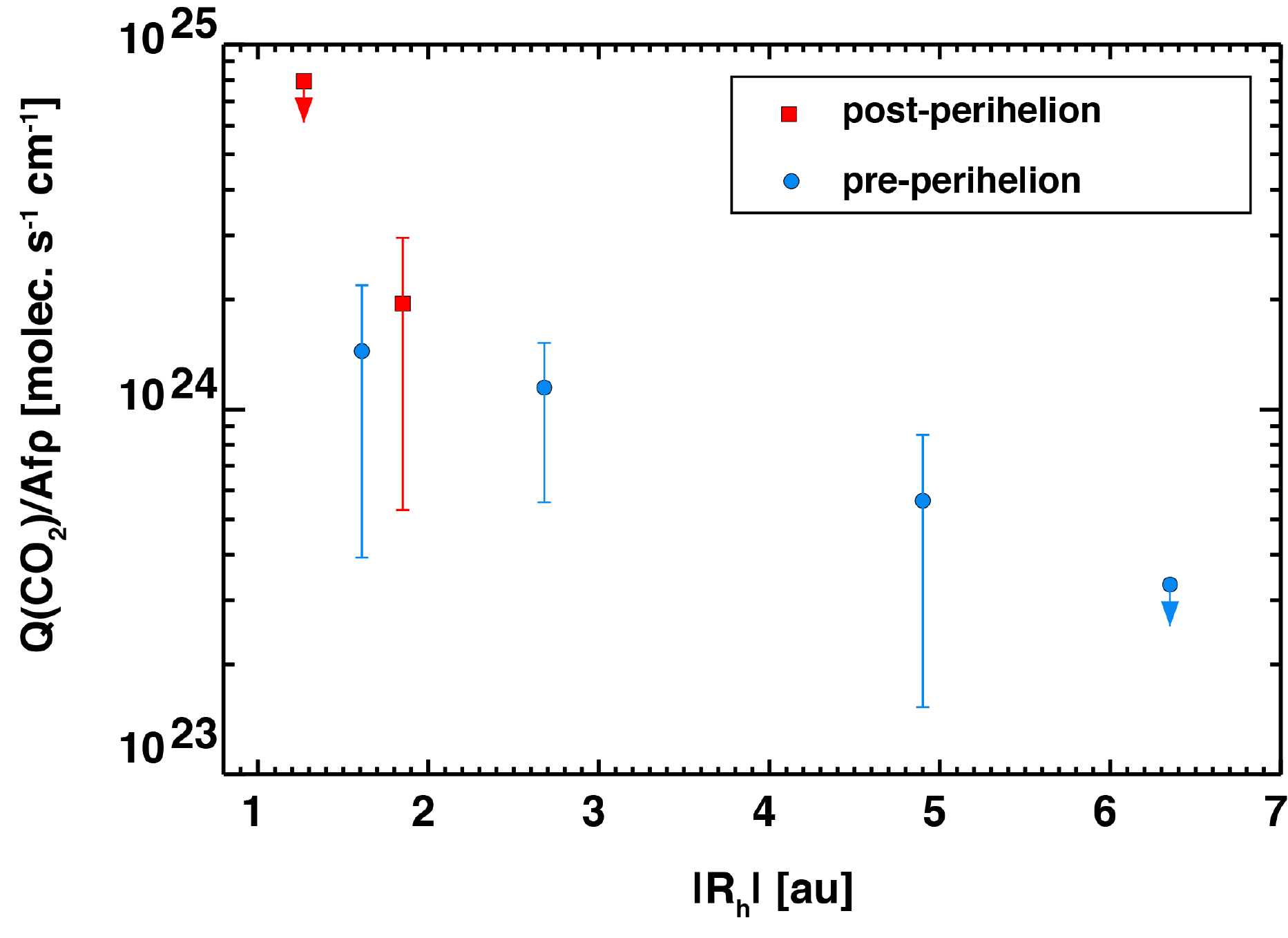}
\caption{\textit{NEOWISE} measurements of Q$_{CO_{2}}$/$Af\rho$ as a function of $\rh$.\label{fig:NEOWISE}}
\end{figure}

The production of water-ice grains throughout the orbit is corroborated by the \textit{NEOWISE} photometry. Sublimation of hyper-volatiles is possibly responsible for dragging water-ice grains into the coma, similarly to what has been observed at comet 103P/Hartley~2.  For comet US10, \coo{} is detected at all visits spanning the range from $-$4.9~au to 1.8~au (Table~\ref{tab_obs}), therefore it is reasonable to expect water ice is produced at all heliocentric distances.  However, the detectability of water ice also depends on the ice-to-dust ratio: a large amount of dust will diminish the ice absorption bands. Considering this scenario, we ratio $Q(\coo)$ with $Af\rho$, and present the results in Fig.~\ref{fig:NEOWISE}.  Here, we find that \coo/$Af\rho$ increases by a factor of 3 from $-$4.9 to 1.8~au. This indicates the lack of water ice in our spectra at 1.3 and 2.3~au is not due to a paucity of \coo{} production or to an abundance of dust production.

We speculate that the observed water-ice physical properties described above may be representative of the interiors of cometary nuclei. This conclusion is supported by the recurrent findings of ``pure'' $\sim$1-\micron~water-ice grains in several comet environments. The icy grains observed in the coma of US10 specifically resemble: (1) those excavated from depths up to 20~m in comet Tempel 1 during the Deep Impact experiment \citep{Sunshine2007}; (2) the icy grains detected in the quiescent sublimation driven coma of comet Hartley 2 by the Deep Impact eXtended Investigation \citep{AHearn2011,Protopapa2014}; (3) ice in the outbursting coma of 67P as seen by the Rosetta mission \citep{Bockelee-Morvan2017, Agarwal2017}; and, (4) ice in cometary comae observed with other ground-based measurements \citep[e.g.,][]{Yang2014}. Furthermore, these icy grains are very different from the $\sim$50-100~\micron~particles seen on the surfaces of both Tempel 1 and 67P \citep{Sunshine2006,Barucci2016}. These similarities and differences suggest that coma ice properties may reflect nucleus interiors rather than processed nucleus surfaces. 

We assert that in comet US10 (1) the sublimation lifetime, and therefore heliocentric distance, controls the physical properties of water ice in the coma; (2) water-ice grains at large heliocentric distance may reflect the comet nucleus interior; and (3) the nucleus is made of, among other components, water-ice grains with particle diameter on the order of $\sim$1~\micron, containing no more than 1\% refractory materials. Is comet US10 atypical? This question can be addressed only by monitoring the properties of the icy grains as insolation varies on the nuclei of several comets, similarly to what has been presented here for comet US10.

The presence of water ice in the coma of comet US10 indicates that it may be a hyperactive comet. However, the definition for a hyperactive comet relies upon an estimate of the effective radius of the nucleus, which is currently lacking for this target.  The comet is at heliocentric distances of 9 to 11~au in 2018.  Imaging of the comet at such great heliocentric distances and beyond may be required in order to measure the nucleus absolute magnitude in the absence of a coma, so that a radius estimate can be made.

\acknowledgements
This work was supported by NASA's SSO grant NNX15AD99G. SP thanks the NASA grant NNX16AC83G for partial funding that supported her work. MSPK acknowledges funding from NASA PMDAP grant NNX13AQ10G.  CEW appreciates support from NASA EW grant NNH15ZDA001N. 

The authors gratefully thank the staff of IRTF for their assistance with this project. This publication also makes use of data products from NEOWISE, which is a project of JPL/Caltech, funded by the Planetary Science Division of NASA. We thank the anonymous referee for valuable suggestions.

This research made use of ephemerides from JPL Horizons \citep{Giorgini1996}.

\facilities{IRTF (SpeX), WISE}
\software{Astropy \citep{astropy}}

\end{document}